\documentclass[
10pt,
showpacs,preprintnumbers,nofootinbib,
 amsmath,amssymb,
 aps,
prl,twocolumn,groupedaddress,superscriptaddress,
showkeys
]{revtex4-1}
\usepackage{graphicx}
\usepackage{dcolumn}
\usepackage{bm}
\usepackage[colorlinks=true,urlcolor=blue,citecolor=blue,breaklinks=true]{hyperref}
\usepackage{color}
\usepackage{epsfig}

\newcommand{\cut}[1]{}

\newcommand{\ket}[1]{\ensuremath{| #1 \rangle}}
\newcommand{\bra}[1]{\ensuremath{\langle #1  |}}

\begin{document}


\title{A Predictive Theory for Elastic Scattering and Recoil of Protons from $^4$He}

\author{Guillaume Hupin}
 \email{ghupin@nd.edu}
 \altaffiliation{Present address:  Department of Physics, University of Notre Dame, Notre Dame, Indiana 46556-5670}
\affiliation{Lawrence Livermore National Laboratory, P.O. Box 808, L-414, Livermore, California 94551, USA}
 \author{Sofia Quaglioni}
 \email{quaglioni1@llnl.gov}
 \affiliation{Lawrence Livermore National Laboratory, P.O. Box 808, L-414, Livermore, California 94551, USA}
 \author{Petr Navr\'atil}
 \email{navratil@triumf.ca}
 \affiliation{TRIUMF, 4004 Wesbrook Mall, Vancouver, British Columbia, V6T 2A3, Canada}

\date{\today}

\begin{abstract}
Low-energy cross sections for elastic scattering and recoil  
of protons from $^4$He nuclei (also known as $\alpha$ particles) are calculated directly by solving the Schr\"odinger equation for five nucleons interacting through accurate two- and three-nucleon forces derived within the framework of chiral effective field theory. Precise knowledge of these processes at various proton backscattering/recoil angles and energies is needed for the ion-beam analysis of numerous materials, from the surface layers of solids, to thin films, to fusion-reactor materials. Indeed, the same elastic scattering process, in two different kinematic configurations, can be used to probe concentrations and depth profiles of either hydrogen or helium.
We compare our results to available experimental data and show that direct calculations with modern nuclear potentials can help to resolve remaining inconsistencies among different data sets and can be used to predict these cross sections when measurements are not available.
\end{abstract}
\maketitle

\paragraph{Introduction.} 
The $^4$He$(p,p)^4$He proton elastic scattering and $^1$H$(\alpha,p)^4$He proton elastic recoil reactions are the leading means for determining the concentrations and depth profiles of, respectively, helium and hydrogen at the surface of materials or in thin films. Such analyses, known among specialists as (non-)Rutherford backscattering spectroscopy and elastic recoil detection analysis, are very important for characterizing the physical, chemical and electrical behavior of materials, for which hydrogen is one of the most common impurities, and for studying the implantation of helium for applications in, e.g., waveguides or fusion energy research~\cite{Paszti1992,Avrahami1996}. To achieve good resolution e.g. in the case of thick samples, measurements are often performed at energies above the Rutherford threshold where the purely Coulomb elastic scattering paradigm does not hold anymore. In  this regime, in which the incident particle energy is on the order of a few MeV per nucleon, nuclear physics becomes the main driver of the scattering process, particularly near low-lying resonances where the cross section can be enhanced by orders of magnitude with respect to the Rutherford rate. Therefore, the availability of accurate reference differential cross sections for a variety of proton/$^4$He incident energies and detection angles are key to the feasibility and quality of these analyses.  

Experimentally the elastic scattering of protons on $^4$He has been studied extensively in the past~ \cite{Freier1949,Kreger1954,Miller1958,Barnard1964,Schwandt1971,Kraus1974}, but only a somewhat limited number of measurements were performed in the energy range of interest for ion-beam analysis, and inconsistencies among different data sets remain~\cite{Baglin1992,Nurmela1997,BogdanovicRadovic2001,Kim2001,Keay2003,Browning2004,Pusa2004}. Consequently, cross sections deduced from {\em R}-matrix analyses of data usually stand as references~\cite{Kraus1974,Dodder1977,Pusa2004,Gurbich2010}. However, there can be discrepancies as large as $10\%$~\cite{Pusa2004} among fits based on different data sets in the critical region near the $3/2^{\texttt-}$ and $1/2^{\texttt-}$ low-lying resonances of $^5$Li. An alternative way of fitting $p$-$^4$He data, based on controlled and systematic effective field theory expansions, was introduced in Ref.~\cite{Higa2008}.  Other theoretical investigations of $p$-$^4$He scattering include microscopic calculations with phenomenological interactions~\cite{Thompson1977,Csoto1997,Kamouni2007} as well as {\em ab initio} calculations based on accurate nucleon-nucleon ($NN$)~\cite{Quaglioni2008} and three-nucleon ($3N$)~\cite{Hupin2013} forces. 
However, both sets of calculations have limited predictive power. The former make use of effective interactions with parameters adjusted to 
reproduce the experimental nucleon-$^4$He  phase shifts~\cite{Kamouni2007},  
and a simplified description of the $^4$He. 
In the latter, an accurate convergence was only achieved for energies above the $^5$Li resonance.
In this Letter we report on  the most complete {\em ab initio} calculation of $p$-$^4$He scattering and provide accurate predictions for 
proton backscattering and recoil cross sections 
at various energies and angles of interest for 
ion-beam applications. 

\paragraph{Formalism.}
We solve the Schr\"odinger equation for $A=5$ interacting nucleons by means of the no-core shell model with continuum (NCSMC)~\cite{Baroni2013}. %
For each channel of total angular momentum, parity and isospin $(J^\pi T)$ we expand the five-nucleon wave function on an overcomplete basis that consists of: $i)$ square-integrable energy eigenstates of the $^5$Li compound system, $\ket{^5 {\rm Li} \, \lambda J^\pi T}$;  and $ii)$ continuous states built from a proton and a $^4$He (or, $\alpha$) nucleus (in a $J_\alpha^{\pi_\alpha}T_\alpha$ eigenstate)  whose centers of mass are separated by the relative coordinate $\vec r_{\alpha,p}$, and that are moving in a $^{2s+1}\ell_J$ partial wave of relative motion,
\begin{align}
\ket{\Phi^{J^\pi T}_{\nu r}} = & 
		\Big[ \!\! \left(
        		\ket{^4 {\rm He} \, \lambda_\alpha J_\alpha^{\pi_\alpha}T_\alpha}\ket{p \, \tfrac12^{\texttt{+}}\tfrac12}
         	\right)^{(sT)} Y_\ell(\hat{r}_{\alpha,p}) \Big]^{(J^{\pi}T)} \nonumber\\ 
	& \times\,\frac{\delta(r{-}r_{\alpha,p})}{rr_{\alpha,p}} \; .
\label{eq:rgm-state}
\end{align}
The resulting NCSMC translational-invariant ansatz is: 
\begin{align}
\ket{\Psi^{J^\pi T}_{A\texttt{=}5}} = &  \sum_\lambda c_\lambda \ket{^5 {\rm Li} \, \lambda J^\pi T} 
+\sum_{\nu}\!\! \int \!\! dr \, r^2 
                 \frac{\gamma_{\nu}(r)}{r}
                 {\mathcal{A}}_\nu \ket{\Phi^{J^\pi T}_{\nu r}} \,.\label{eq:ansatz}
\end{align}
The $^4$He and $^5$Li states of Eqs.~(\ref{eq:rgm-state}) and (\ref{eq:ansatz}), identified respectively by the energy labels $\lambda_\alpha$ and $\lambda$, are  
antisymmetric under exchange of internal nucleons. They are obtained ahead of time by means of the {\em ab initio} no-core shell model~\cite{Navratil2000a} through the diagonalization of their respective microscopic Hamiltonians in finite bases constructed from many-body harmonic oscillator (HO) wave functions with up to $N_{\rm max}$ HO quanta and frequency $\hbar\Omega$. The index $\nu$ collects the quantum numbers 
$\{^4 {\rm He} \; \lambda_\alpha J_\alpha^{\pi_\alpha}T_\alpha; p \; \tfrac12^{\texttt{+}}\tfrac12; s\ell\}$ associated with the continuous basis states of Eq.~(\ref{eq:rgm-state}), 
and the operator ${\mathcal{A}}_\nu= \tfrac1{\sqrt{5}}(1 - \sum_{i=1}^{4} P_{i,5})$, with $P_{i,5}$ the permutation  
between a nucleon belonging to the $^4$He nucleus and the proton, ensures the full antisymmetrization of the five-nucleon system.  The discrete coefficients, $c_\lambda$, and the continuous
amplitudes of relative motion, $\gamma_{\nu}(r) = ({\mathcal N}^{-1/2}\chi)_\nu(r)$, are the unknowns of the problem and are obtained as solutions, in the interaction region, of the coupled equations \begin{eqnarray}
\left(
\begin{array}{cc}
        H_{^5 {\rm Li}} & \bar{h} \\  \bar{h}  & \overline{\mathcal{H}} 
\end{array}
\right) \left(
\begin{array}{c}
	c \\ {\chi}
\end{array}
\right)  =   E \left(
\begin{array}{cc}
        I_{^5 {\rm Li}} & \bar{g} \\  \bar{g}  & {\mathcal I} 
\end{array}
\right) \left(
\begin{array}{c}
	c \\  {\chi}
\end{array}
\right). \label{eq:NCSMC-eq}
\end{eqnarray}
Here, $E$ denotes the total energy of the system and the two by two block-matrices on the left- and right-hand side of the equation represent, respectively, the NCSMC Hamiltonian and norm 
kernels.  In the upper diagonal block one can recognize the matrix elements of the Hamiltonian $H$ (identity $I$) over the discrete $^5$Li states, $(H_{^5 {\rm Li}})_{\lambda\lambda'}\!=\!\delta_{\lambda \lambda'}E_\lambda$ [$(I_{^5 {\rm Li}})_{\lambda\lambda'}\!=\!\delta_{\lambda\lambda^\prime}$]. Similarly, those over the orthonormalized $p$-$^4$He portion of the basis, $\overline{ \mathcal{H}}_{\nu\nu^\prime}(r,r^\prime)\!=\! (\mathcal{N}^{-1/2}\mathcal{H}\mathcal{N}^{-1/2})_{\nu\nu^\prime}(r,r^\prime)$ [${ \mathcal{I}}_{\nu\nu^\prime} (r,r^\prime)\!=\! \delta_{\nu\nu^\prime}\delta(r-r^\prime)/(r r^\prime)$], which are obtained from ${\mathcal N}_{\nu\nu^\prime}(r,r^\prime)\!=\!\bra{\Phi^{J^\pi T}_{\nu r}} {\mathcal A}_\nu{\mathcal A}_{\nu^\prime}\ket{\Phi^{J^\pi T}_{\nu^\prime r^\prime}}$ and ${\mathcal H}_{\nu\nu^\prime}(r,r^\prime)\!=\!\bra{\Phi^{J^\pi T}_{\nu r}} {\mathcal A}_\nu H {\mathcal A}_{\nu^\prime}\ket{\Phi^{J^\pi T}_{\nu^\prime r^\prime}}$, appear in the lower diagonal block. The 
couplings between the two sectors of the basis are described by the overlap, $\bar{g}_{\lambda \nu}(r)\!=\!(g{\mathcal N}^{-1/2})_{\lambda \nu}(r)$, and Hamiltonian, $\bar{h}_{\lambda \nu}(r)\!=\!(h{\mathcal N}^{-1/2})_{\lambda \nu}(r)$, form factors, with $g_{\lambda \nu}(r)\!=\!\bra{^5 {\rm Li} \, \lambda J^\pi T} {\mathcal A}_{\nu}\ket{\Phi^{J^\pi T}_{\nu r}}$ and $h_{\lambda \nu}(r)\!=\!\bra{^5 {\rm Li} \, \lambda J^\pi T}H {\mathcal A}_{\nu}\ket{\Phi^{J^\pi T}_{\nu r}}$. 
The scattering matrix (and from it any scattering observable) is then obtained by matching the solutions of Eq.~(\ref{eq:NCSMC-eq}) with the known asymptotic behavior of the wave function at large distances by means of the microscopic $R$-matrix method~\cite{Hesse1998,Hesse2002}.

\paragraph{Results.}
\begin{figure}[t]
\flushleft
\includegraphics*[width=0.85\linewidth]{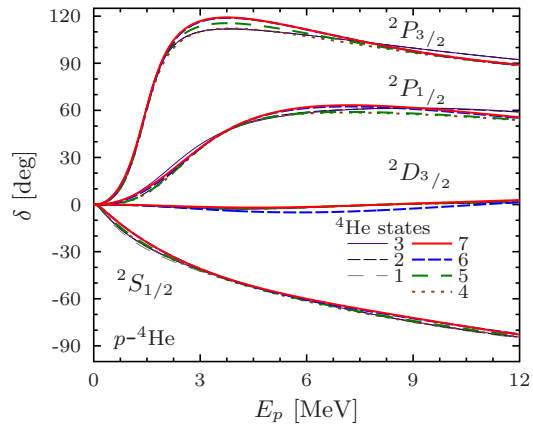}
\caption{(Color online) Calculated $p$-$^4$He phase shifts at $N_{\rm max}=13$ and $\hbar\Omega=20$ MeV obtained with up to fourteen states of the compound $^5$Li nucleus as a function of the number of $^4$He states included in the calculation. The solid red lines represent our most complete results.   
All values in this and the subsequent figures are in the laboratory frame.} \label{fig:CV-ExcitedStates}
\end{figure}
Different from Refs.~\cite{Baroni2013}, where the NCSMC 
was introduced and applied to the description of the unbound $^7$He nucleus starting from an accurate $NN$ potential, here we 
employ  
this approach for the first time with an Hamiltonian that also includes in addition explicit $3N$ forces.  
This is, from an {\em ab initio} standpoint, necessary to obtain a truly accurate and quantitative description of the scattering process~\cite{Nollett2007,Viviani2013}. In particular, we adopt an Hamiltonian based on the chiral N$^3$LO  
$NN$ interaction of Ref.~\cite{Entem2003} and N$^2$LO  
$3N$ force of Ref.~\cite{Navratil2007}, constrained to provide an accurate description of the $A=2$ and $3$~\cite{Gazit2009} systems and unitarily softened via the similarity-renormalization-group (SRG) method~\cite{Glazek1993,Wegner1994,Bogner2007,Hergert2007,Jurgenson2009} to minimize the influence of momenta higher than 
2 fm$^{-1}$.

An {\em ab initio} investigation of elastic scattering of protons on $^4$He using the present Hamiltonian was recently obtained  
within the continuous sector only of the model space considered here [corresponding to the second term in the right-hand side of Eq.~(\ref{eq:ansatz})], i.e.\ by solving $\overline{ \mathcal{H}}\chi=E\chi$~\cite{Hupin2013}. 
There, a careful analysis of the computed scattering phase shifts showed that  
independence 
with respect to the parameters characterizing the 
HO basis is approached at $N_{\rm max}=13$ (currently our computational limit) and $\hbar\Omega=20$ MeV, and 
that small variations of  
the SRG momentum scale  
around the value chosen here ($\Lambda=2.0$ fm$^{-1}$) do not lead to significant differences in the results.
By far the largest variation in the obtained phase shifts was observed as a function of the number of states used to describe the helium nucleus, 
particularly in the $^2P_{3/2}$ and $^2P_{1/2}$ partial waves, 
where even the inclusion of up to the first seven ($J_\alpha^{\pi_\alpha} T_\alpha =$ 0$_1^{\texttt+}$0, 0$_2^{\texttt+}$0, 0$^{\texttt-}$0, 2$^{\texttt-}$0, 2$^{\texttt-}$1, 1$^{\texttt-}$1 and 1$^{\texttt-}$0) $^4$He eigenstates proved to be insufficient for the accurate description of the resonances below $E_p\sim5$ MeV proton incident energy (see Fig.\ 10 of Ref.~\cite{Hupin2013}).

Rather than adding higher helium excitations, which would lead to a computationally unbearable problem, here we augment the model space adopted in Ref.~\cite{Hupin2013} by coupling the first fourteen (of which three $3/2^{\texttt-}$ and two $1/2^{\texttt-}$) square-integrable eigenstates of the $^5$Li compound nucleus. As illustrated in Fig.~\ref{fig:CV-ExcitedStates}, and previously demonstrated with a two-body Hamiltonian for neutron-$^6$He scattering~\cite{Baroni2013}, this substantially mitigates the dependence on the number of eigenstates of the target so that even a model space including only the ground state (g.s.)\ of $^4$He is already sufficient to provide a reasonable description of the significant elastic scattering phase shifts. 
Still, to reach the high accuracy we seek in the present work  higher helium excitations cannot be neglected. This is because in spite of the correlations added by the $^5$Li compound states, 
the $J_\alpha^{\pi_\alpha} T_\alpha = 0^{\texttt-}0$, $2^{\texttt-}0$, $2^{\texttt-}1$ and $1^{\texttt-}1$ (respectively the third, fourth, fifth and sixth) states do play a role, particularly in determining the $3/2^{\texttt-}$ and $1/2^{\texttt-}$ resonance energies and widths. \begin{figure}[t]
\centering
\includegraphics*[width=.85\linewidth]{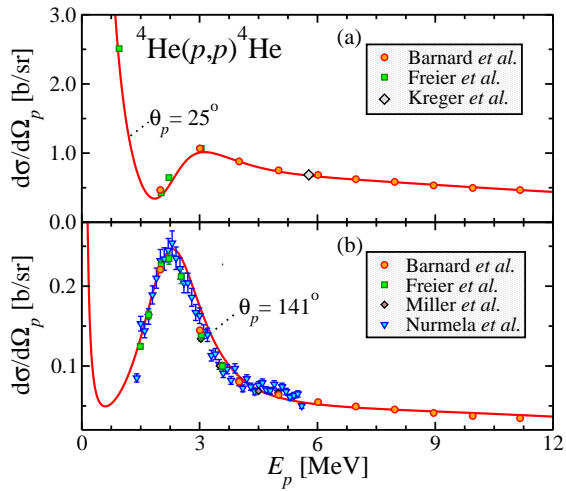}
\caption{(Color online) Computed (solid red lines) $^4$He$(p,p)^4$H angular differential cross section at forward scattering angle $\theta_p=25^\circ$ (a) and backscattering angle $\theta_p=141^\circ$ (b) as a function of the proton incident energy compared with measurements (symbols) from Refs.~\cite{Freier1949,Kreger1954,Miller1958,Barnard1964,Nurmela1997}. The calculation corresponds to the most complete results of Fig.~\ref{fig:CV-ExcitedStates}.} \label{fig:scattering-0to12MeV}
\end{figure}

In Fig.~\ref{fig:scattering-0to12MeV} our most complete results (including the first seven low-lying states of $^4$He) for the $^4$He$(p,p)^4$He angular differential cross section at the laboratory proton-scattering angles of $\theta_p=25^\circ$ and $141^\circ$ are compared to measurements in the range of incident energies up to $12$ MeV~\cite{Freier1949,Kreger1954,Miller1958,Barnard1964,Nurmela1997} . The agreement with data is excellent both at forward and backward angles. 
The high energy tail of the cross section was already well described within the more limited 
model space of Ref.~\cite{Hupin2013}. 
The effect of the additional  
$^5$Li states,  
included in the present calculation, is essentially confined around their 
eigenenergies. The first $3/2^{\texttt-}$ and $1/2^{\texttt-}$ states play the largest role, substantially improving the agreement with experiment at lower energies. Indeed, we see in Fig.~\ref{fig:scattering-0to12MeV} that the calculated differential cross section lies within the experimental error bars in the peak region dominated by the resonances, though the width of the peak is somewhat overestimated. 
\begin{table}[t]
\caption{Centroids $E_R$, obtained as the values of the kinetic energy in the center of mass for which the first derivative $\delta^\prime(E_{\rm kin})$ of the phase shift is maximal~\cite{Csoto1997}, and widths $\Gamma=2/\delta^\prime(E_R)$ of the $^5$Li ground and first excited states. The $R$-matrix results are taken from Ref.~\cite{Csoto1997} and correspond to the evaluation of Ref.~\cite{Tilley2002}. Units are in MeV.
\label{table:resonance}}
\begin{ruledtabular}
\renewcommand{\arraystretch}{1.3}
\begin{tabular}{c  c  c   c  c}
& \multicolumn{2}{c}{$R$-matrix} & \multicolumn{2}{c}{Present results}\\
$J^{\pi}$&$E_R$\,&$\Gamma$\,&$E_R$\,&$\Gamma$\,\\
\hline
${3/2}^-$	&	1.67	&	1.37	&	1.77(1)      & 1.70(5)\phantom{0}	\\
${1/2}^-$	&	3.35	&	9.40	&	3.11(2)       & 7.90(50)	
\end{tabular}
\end{ruledtabular}
\end{table}

In Table~\ref{table:resonance}, we compare the present results for the centroids and widths of the $^5$Li ground and first excited states to  
those from an extended $R$-matrix analysis of data~\cite{Csoto1997}.\begin{figure}[b]
\centering
\includegraphics*[width=.85\linewidth]{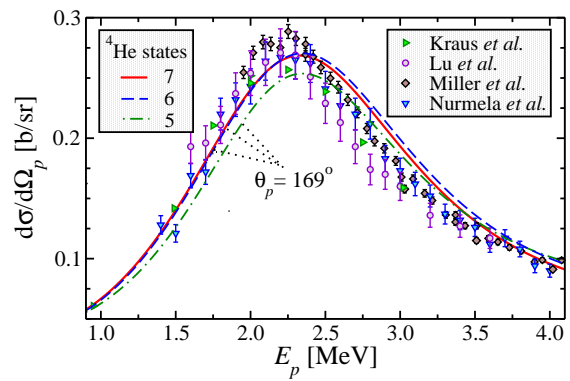}
\caption{(Color online) Same as Fig.~\ref{fig:scattering-0to12MeV} but at the backscattering angle of $\theta_p=169^\circ$ and in the range of proton incident energies near the $^5$Li resonances. Calculations including 5 and 6 $^4$He states are shown in addition to the most complete results. Experimental data are from Refs.~\cite{Miller1958,Kraus1974,Lu2009,Nurmela1997}.} \label{fig:backscattering}
\end{figure}
\begin{figure*}
\begin{minipage}[c]{.48\linewidth}
\centering
\includegraphics*[width=.85\linewidth]{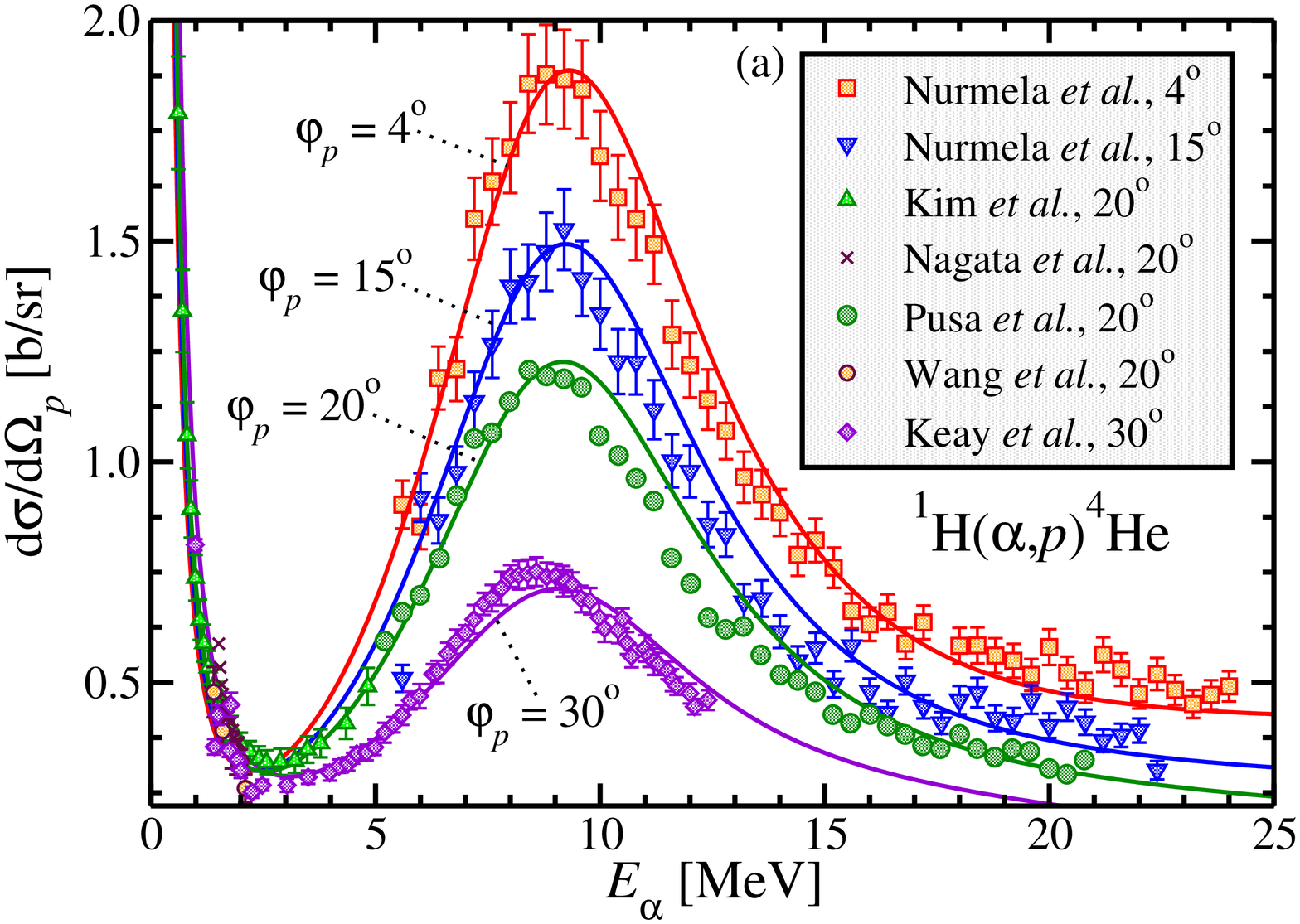}
\end{minipage}
\hfill
\begin{minipage}[c]{.48\linewidth}
\centering
\includegraphics*[width=.85\linewidth]{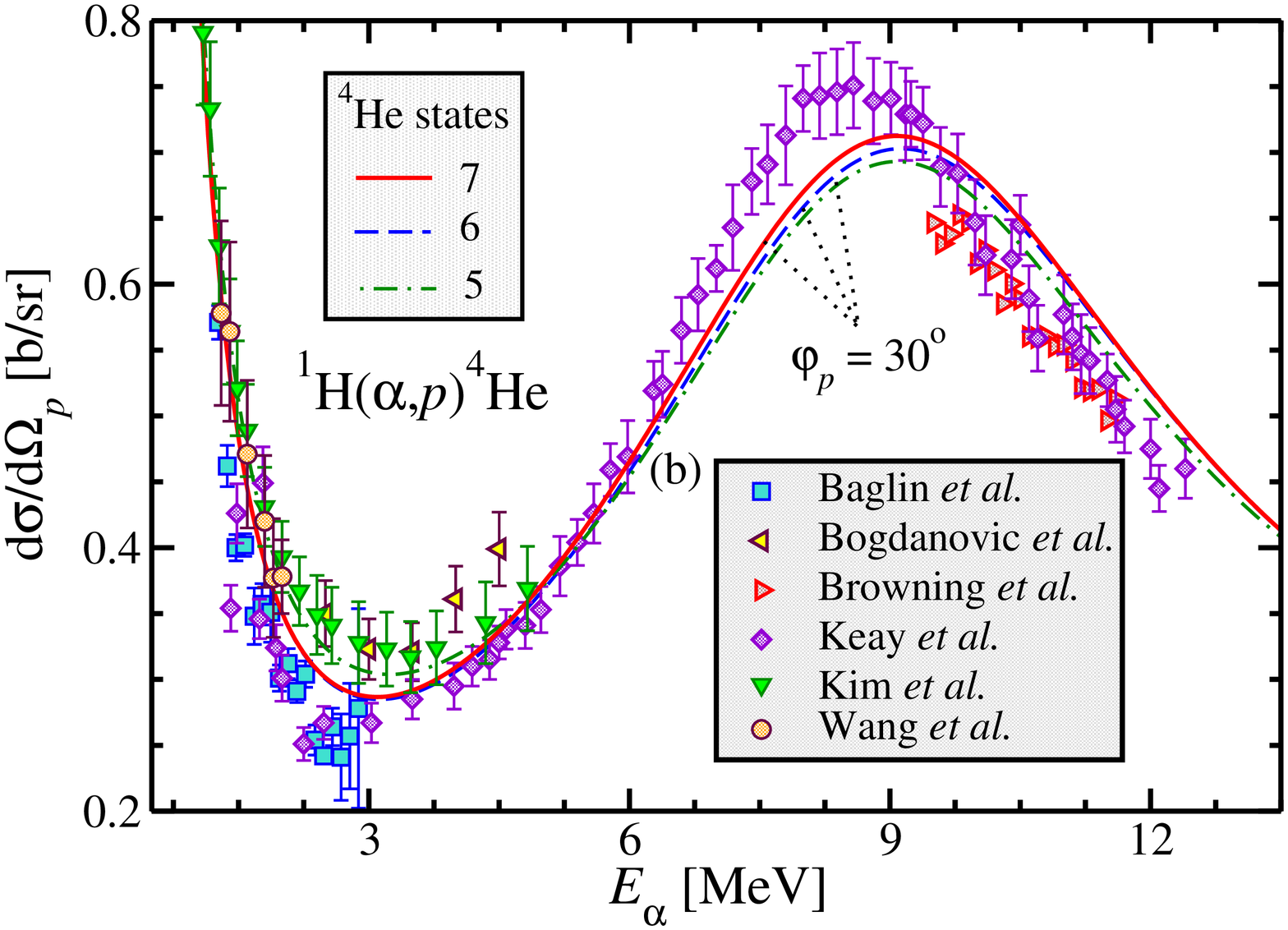}
\end{minipage}
\caption{(Color online) Computed (lines) $^1$H$(\alpha,p)^4$He angular differential cross section at the proton recoil angles $\varphi_p=4^\circ, 16^\circ, 20^\circ$, and $30^\circ$ as a function of the incident $^4$He energy compared with data (symbols) from Refs.~\cite{Nagata1985,Baglin1992,Wang1986,Nurmela1997,BogdanovicRadovic2001,Kim2001,Keay2003,Pusa2004,Browning2004}. Panel (b) focuses on the proton recoil angle $\varphi_p=30^\circ$, and shows, in addition to the most complete results, calculations including 5 and 6 $^4$He states.} \label{fig:proton-recoil}
\end{figure*}
The resonance 
positions are in fairly good agreement. The largest deviation occurs for the $1/2^{\texttt-}$ state, which is $240$ keV below the energy reported in Ref.~\cite{Csoto1997}.  However we find larger differences for the widths, particularly for 
the $^5$Li g.s., 
which is $24\%$  
broader than in the $R$-matrix analysis. 
The computed widths, particularly that of the $1/2^{\textit{-}}$ resonance, present the largest uncertainty in terms of number of $^4$He states included in the calculation (indicated in parenthesis). 
In Fig.~\ref{fig:backscattering}, we zoom  
to energies near the resonances at the proton scattering angle of $169^\circ$, of interest for non-Rutherford backscattering spectroscopy, where the $R$-matrix analysis of Ref.~\cite{Dodder1977} leads to an overestimation of the cross section and triggered the search for new fitting parameters~\cite{Pusa2004}. Except for the $2.4$ MeV$\le E_p\le 3.5$ MeV energy interval, where there is a minor disagreement with experiment in line with our previous discussion, the computed cross section is in overall satisfactory agreement with data and shows that the present theory could provide accurate guidance for ion beam analyses at energies/angles where measurements are not available. The theoretical uncertainty 
associated with the treatment of the helium excitations can be estimated from Fig~\ref{fig:backscattering}, by studying the convergence of the  
cross section with respect to the last three $^4$He  
states included in the calculation. The three curves are all within 5\% one from another 
and differences between the results with 6 and 7  
states are minimal. This and the results of Table~\ref{table:resonance} earlier point to remaining deficiencies in 
 the nuclear interaction (and in particular $3N$ force) used in this work. In fact, refinement of the chiral $3N$ force (which affects the spin-orbit splitting between $^{2}P_{3/2}$ and $^{2}P_{1/2}$ phase shifts) is a current topic of interest in nuclear physics~\cite{Krebs2012,Kalantar-Nayestanaki2012,Krebs2013}.    

\begin{figure}[b]
\includegraphics*[width=.85\linewidth]{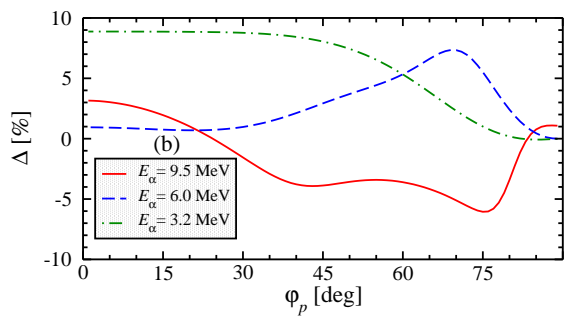}
\caption{(Color online) Relative difference (in percent)  between the calculated elastic recoil cross section at $N_{\rm max}=13$ and 11 as a function of the proton angle $\varphi_p$ for the helium incident energies $E_\alpha=3.2$, $6.0$ and  $9.5$ MeV. Only the first two $^5$Li states are accounted for in this study.} \label{fig:uncert}
\end{figure}
Another kinematic setting of interest is the elastic recoil of protons at forward angles by incident $^4$He nuclei. 
In Fig.~\ref{fig:proton-recoil}(a), the computed $^1$H$(\alpha,p)^4$He angular differential cross section at the proton recoil angles $\varphi_p=4^\circ$, $15^\circ$, $20^\circ$ and $30^\circ$ is compared to various data sets over a wide range of helium incident energies, $E_\alpha$.  For all four angles the agreement with experiment is excellent close to the Rutherford threshold (particularly at the base of the cross section) and above $E_\alpha\sim13$ MeV, but  once again deteriorates at intermediate energies due to the overestimated  width of the $3/2^{\texttt-}$ resonance. 
In Fig.~\ref{fig:proton-recoil}(b), we concentrate on the well-studied proton recoil angle of $\sim30^\circ$. 
In the dip near $E_\alpha=3$ MeV, where the cross section is fairly insensitive to the recoil angle, measurements differ up to 40\%. On the contrary our results, which lie in between the data of Baglin {\em et al}.~\cite{Baglin1992}  
and those of 
Kim {\em et al}.~\cite{Kim2001},  are very stable with respect to the number of helium states included in the calculation at this energy, and a study of the uncertainty associated with the size of the HO basis $N_{\rm max}$, shown in Fig.~\ref{fig:uncert}, indicates that  they are accurate to less than 10\%. 
However, different from the trend observed at the smaller recoil angles, our calculation here underestimates measurements in the peak region. The extent of this deviation goes beyond the numerical error due to our finite model space and 
is likely to be associated with the remaining uncertainties in the nuclear Hamiltonian.

\paragraph{Conclusions.}
We presented the most advanced {\em ab initio} calculation of  $p$-$^4$He elastic scattering and provided accurate predictions for proton backscattering and recoil cross sections at various energies and angles of interests to ion beam spectroscopy.  Our statistical error, due to the finite size of the model space, is within 9\%. This is of the same order as 
experimental uncertainties. An in depth investigation of the systematic error associated with the nuclear Hamiltonian is beyond the scope of the present work. However, we found evidence that the present interaction leads to an overestimation of the width of the $^5$Li g.s.\ resonance as well as to a somewhat insufficient splitting between this and the $1/2^{\texttt-}$ excited state. 
With the ability to further reduce and control the theoretical uncertainties spurred by the development of optimized nuclear interactions and exascale computing capabilities, the direct solution of the Schr\"odinger equation is poised to become a competitive approach to provide guidance for applications using light-nucleus cross sections. 

\begin{acknowledgments}
Computing support for this work came from the LLNL institutional Computing Grand Challenge
program. Prepared in part by LLNL under Contract DE-AC52-07NA27344. This material is based upon work supported by the U.S.\ Department of Energy, Office of Science, Office of Nuclear Physics, under 
Work Proposal Number SCW1158, 
and by the NSERC Grant Number 401945-2011. TRIUMF receives funding via a contribution through the 
Canadian National Research Council.
\end{acknowledgments}

\bibliographystyle{apsrev4-1}

%

\end{document}